\documentclass[%
 reprint,
 amsmath,amssymb,
 aps,
]{revtex4-2}

\usepackage{hyperref}
\usepackage{placeins}
\usepackage{graphicx}
\usepackage{dcolumn}
\usepackage{bm}

\begin{document}

\preprint{APS/123-QED}

\title{ A background for thermal photons in heavy ion collisions  \\}

\author{Satya Ranjan Nayak$^{1}$}
 \email{satyanayak@bhu.ac.in}
\author{Gauri Devi$^{1}$}
 \email{gauri1995devi@gmail.com}
\author{B. K. Singh$^{1,2}$}%
 \email{bksingh@bhu.ac.in}
 \email{director@iiitdmj.ac.in}
\affiliation{$^{1}$Department of Physics, Institute of Science,\\ Banaras Hindu University (BHU), Varanasi, 221005, INDIA. \\
$^{2}$Discipline of Natural Sciences, PDPM Indian Institute of Information Technology Design \& Manufacturing, Jabalpur 482005, INDIA.}
\date{\today}

\date{\today}
\begin{abstract}
In this work, we present the transverse momentum spectra of prompt and decay photons in Au-Au collisions for $\sqrt{s_{NN}}=$ 200 GeV, 62.4 GeV, 39 GeV, and 27 GeV. The major sources of the photons in Angantyr include hard processes, Parton showers, and resonance decay. The multiparton interactions and hadronic rescatterings significantly increase the photon yield. The model shows a good match with the available experimental data at high $p_T$. The difference in yield at low $p_T $ suggests that Quark Gluon Plasma of $T_{eff}$ = 0.167 GeV/c in central Au-Au collision at 200 GeV is formed, the new effective temperature is less than the ones extracted without removing background photons. At low $p_T$ the decay photon spectra scales with $(\frac{dN_{ch}}{d\eta})^{1.25}$, the scaling is independent of collision energy and system size. The scaling no longer holds at high $p_T$ and the spectra become beam energy dependent. The scaled $p_T$ spectra of p-p and d-Au collisions show an opposite trend at high $p_T$, their scaled yield is greater than the Au-Au collision at the same energy.

\end{abstract}

\maketitle


\section{Introduction}

The heavy ion collision experiments produce a hot dense state of de-confined partons called the quark-gluon plasma \cite{Collins:1974ky,Cabibbo:1975ig,Chapline:1976gy}. Such a system can be identified by it's signatures \cite{singh1993signals} like suppressed high $p_T$ spectra \cite{PHENIX:2001hpc}, elliptic flow \cite{Wiedemann:1997cr}, strangeness abundance \cite{Rafelski:1982pu,Rafelski:1982ii}, $j/\psi$ suppression \cite{Wong:1997rm}, etc. The spectra of direct photons are an excellent way of studying such a system. The photons have a large mean free path and rarely interact with the quark-gluon plasma. Hence, they carry information about all stages of collision. The direct photons have been observed in different experiments at a range of collision energies \cite{WA98:2000vxl,PHENIX:2008uif}.

The major sources of photons in the A-A or p-p collision systems include hard scattering, secondary hard scattering, soft processes, initial state and final state radiations, and resonance decays \cite{Sjostrand:2004ef}. The weak showers can also be a source of photons but their effects are very small. Other exotic processes involving one extra dimension can also produce photons via graviton exchange. However, such processes are beyond the scope of this work \cite{Azuelos:2004ux}. The major distinction between a p-p collision and an A-A collision is the existence of QGP which itself emits photons, such photons are called thermal photons. Many theoretical approaches have been made to study direct photons like the elliptic fireball, thermalized glasma, hydrodynamic simulations of the “fireball” evolution, etc \cite{vanHees:2011vb,vanHees:2014ida,Dion:2011pp,McLerran:2014hza}. The recent experiments under the RHIC Beam Energy Scan (BES) measured the direct photon spectra for lower collision energies like $\sqrt{s_{NN}}=$ 62.4 GeV and 39.0 GeV. The measurements found the excess of direct photons which are thermal. Thermal photons can also be found in smaller collision systems like d-Au and 3He-Au \cite{PHENIX:2018hho,PHENIX:2023kax}. These systems show QGP-like signals such as jet quenching and elliptic flow but fail to produce signals like baryon enhancement. The existence of thermal photons in these systems would imply the existence of QGP.  However, effectively distinguishing thermal photons from other sources is a difficult problem due to the contribution of a large number of processes. The major aim of this study is to find the contributions from different processes in photon spectra which will help to distinguish thermal photons from others. We are using PYTHIA8 Angantyr for this study \cite{Bierlich:2018xfw,Bierlich:2022pfr}. The Angantyr model combines individual NN sub-collisions from PYTHIA8 based on the FRITIOF model. The model does not consider a hot thermalized medium (QGP). Hence, it can give a clear picture of the non-collective photon spectra in various collisions. A detailed description of the model can be found in the next section.

This letter is organized as follows, in section II we have briefly described the Angantyr model. In section III A we have shown different sources of photons and the effect of parton showers, Mulit-Parton interactions, and rescattering. In section III B we have discussed the prompt photon production in smaller collision systems. Finally, in section IV we have summarized our work.

\section{Angantyr}

PYTHIA8 recently developed the Angantyr algorithm to simulate nucleus-nucleus collisions and proton-nucleus collisions \cite{Bierlich:2018xfw,Bierlich:2022pfr}. The FRITIOF program and the wounded nucleon model inspired the model with added effects from partonic hard scattering \cite{Galoyan:2002rdi}. The positions and the number of the interacting nucleons and binary nucleon-nucleon collisions are calculated using Glauber-model-based eikonal approximation in impact-parameter space and Gribov’s corrections on color fluctuations are implemented to include diffractive excitation \cite{Heiselberg:1991is,Alvioli:2014eda}, which appears due to the fluctuations in the nucleon substructure \cite{Bierlich:2018xfw}. The effects of such fluctuations can be found in the corresponding paper \cite{ATLAS:2015hkr}. The hard partonic sub-collisions, normalized by nucleon-nucleon sub-collisions, play a major role at high energies. The model separates the individual NN sub-collisions as an absorptive, wounded target, wounded projectile, double diffractive, and elastic interactions. The absorptive interactions are treated as pp-like Non-Diffractive interactions. The scenario where a single target or projectile nucleon is wounded is treated similarly to Single-Diffractive interaction, the double diffractive and elastic sub-collisions are treated accordingly. All subevents are then combined to simulate a p-A or A-A event. This can give a better and more accurate description of a heavy ion event than a scaled p-p event\cite{Lonnblad:2019dyj}.


\begin{table*}
\begin{tabular}{c c c c c c c} 
 \hline
 System & $\sqrt{s_{NN}}$ GeV & Centrality \% & $p_{T0}^{ref}$ GeV/c& $\frac{dN_{ch}}{d\eta}|_{\eta=0}$ & data (PHENIX) & model/data\\ 
     
 \hline
 Au-Au & 	200  &(0-20)\% & 2.44 & 519.5 & 519.02 $\pm$26.25 & 1.000\\ 

Au-Au & 62.4   &(0-20)\% & 2.68 & 341.4  & 341.175	$\pm$ 29.325 & 1.001 \\

Au-Au & 39   &(0-20)\% & 2.85 & 278.4 & 278.0  $\pm$ 24.05	& 1.001 \\

Au-Au & 27   &(0-20)\% & 2.88 & 242.05	& 241.0 $\pm$ 21.5 & 1.004\\

d-Au & 200   &(0-20)\% & 3.09 & 16.85	& 16.8 $\pm$ 1.17 &1.002 \\

3He-Au & 200   &(0-20)\% & 3.05 & 23.0 & 22.9$\pm$1.6 & 1.004  \\

 \hline
\end{tabular}
\caption{The values of $p^{ref}_{T0}$ and multiplicity for various collision energies, centrality, and system.
}
\label{Table1}
\end{table*}

\begin{figure*}
\includegraphics[width=.58\textwidth]{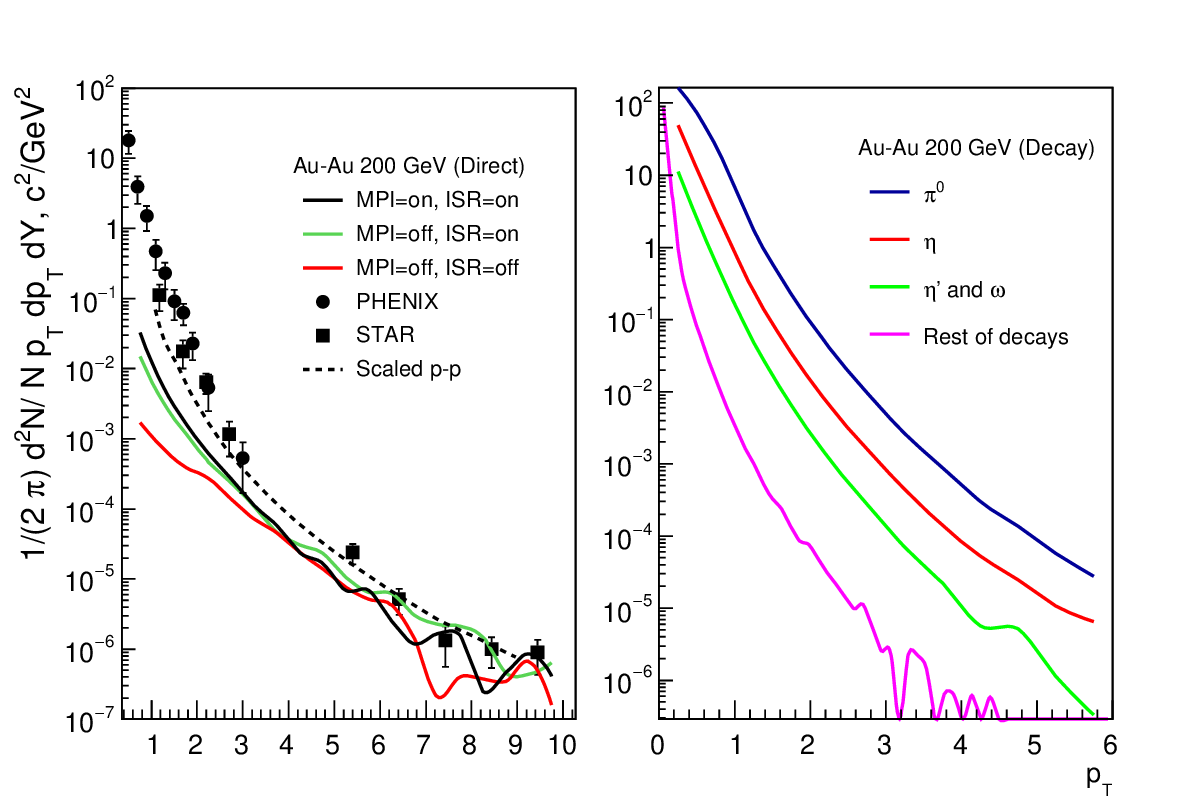}
\caption{\label{fig:2} The invariant yield of direct and decay photons for most central Au-Au collisions at 200 GeV. Lines show the Angantyr results and markers represent experimental data \cite{STAR:2016use,PHENIX:2022rsx}. }
\end{figure*}

\begin{figure}[h]
\includegraphics[width=.5\textwidth]{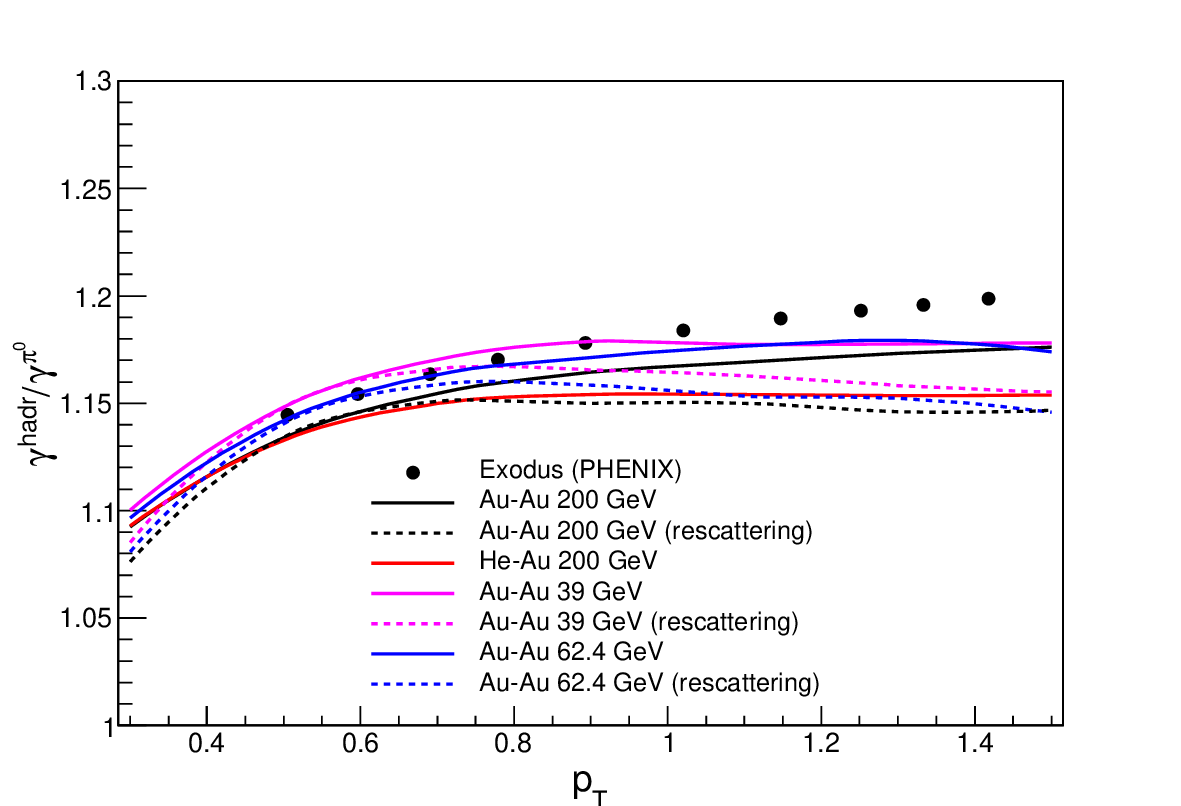}
\caption{\label{fig:3} The $\gamma ^{hadrons}/\gamma ^{\pi ^0}$ ratios for different collision systems and energies details are mentioned in the legend. }
\end{figure}
\begin{figure}[h]
\includegraphics[width=.5\textwidth]{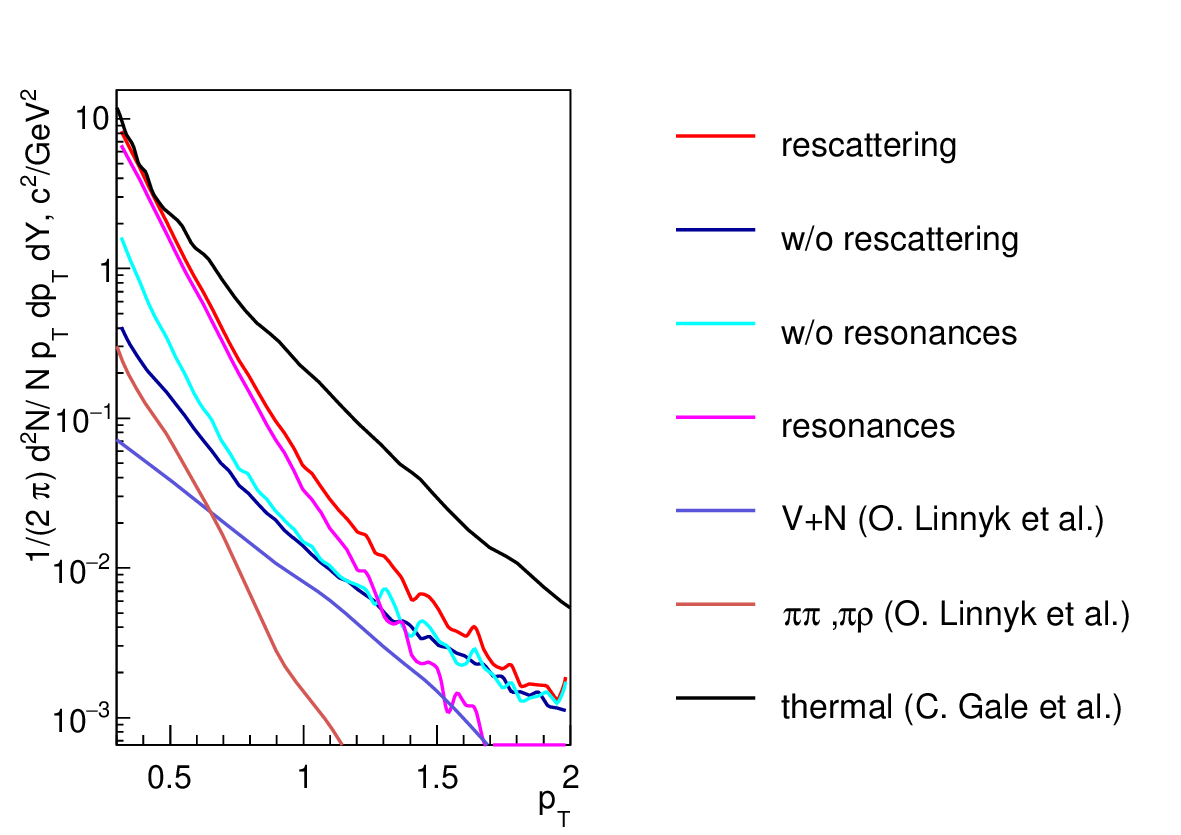}
\caption{\label{fig:3} The photon spectra in Au-Au collisions at 200 GeV with rescattering. }
\end{figure}

\begin{figure}[h]
\includegraphics[width=.5\textwidth]{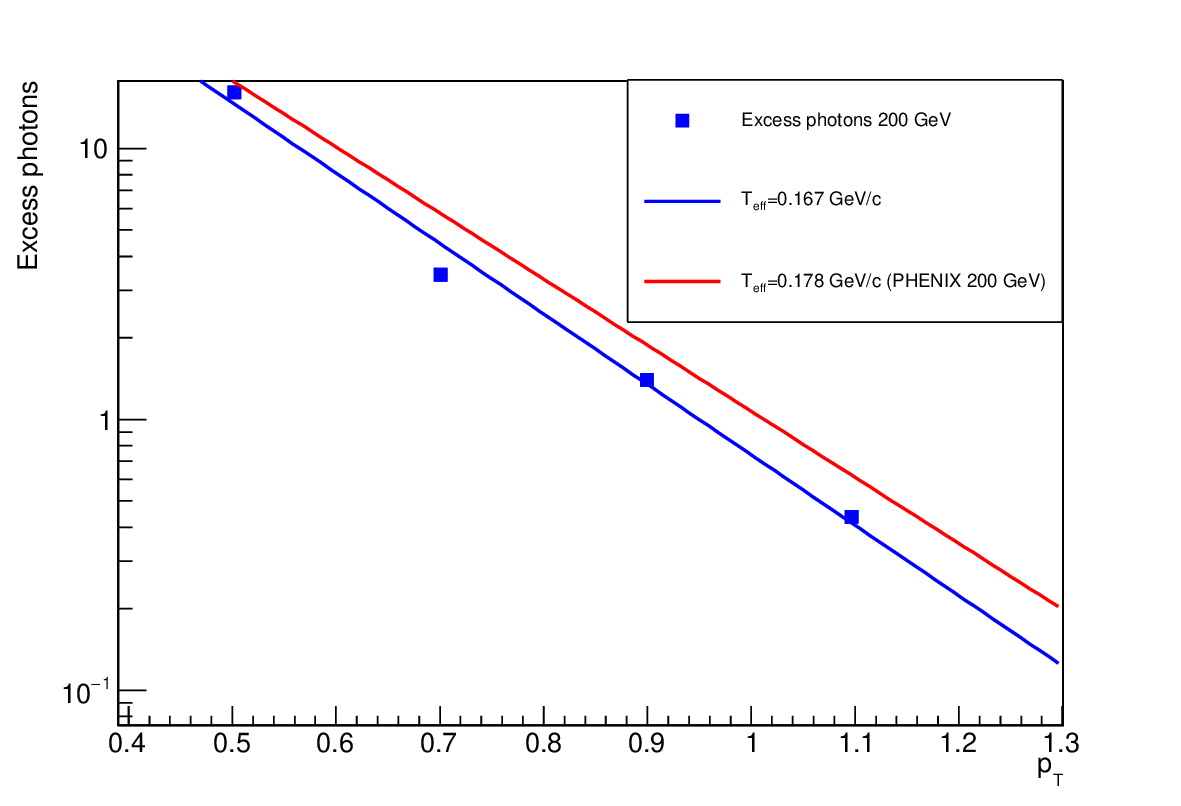}
\caption{\label{fig:3} The difference between experimental data and simulated prompt photon yields are shown with markers. Lines show the exponential fit with the effective temperature mentioned in the legend. }
\end{figure}
\begin{figure}[h]
\includegraphics[width=.5\textwidth]{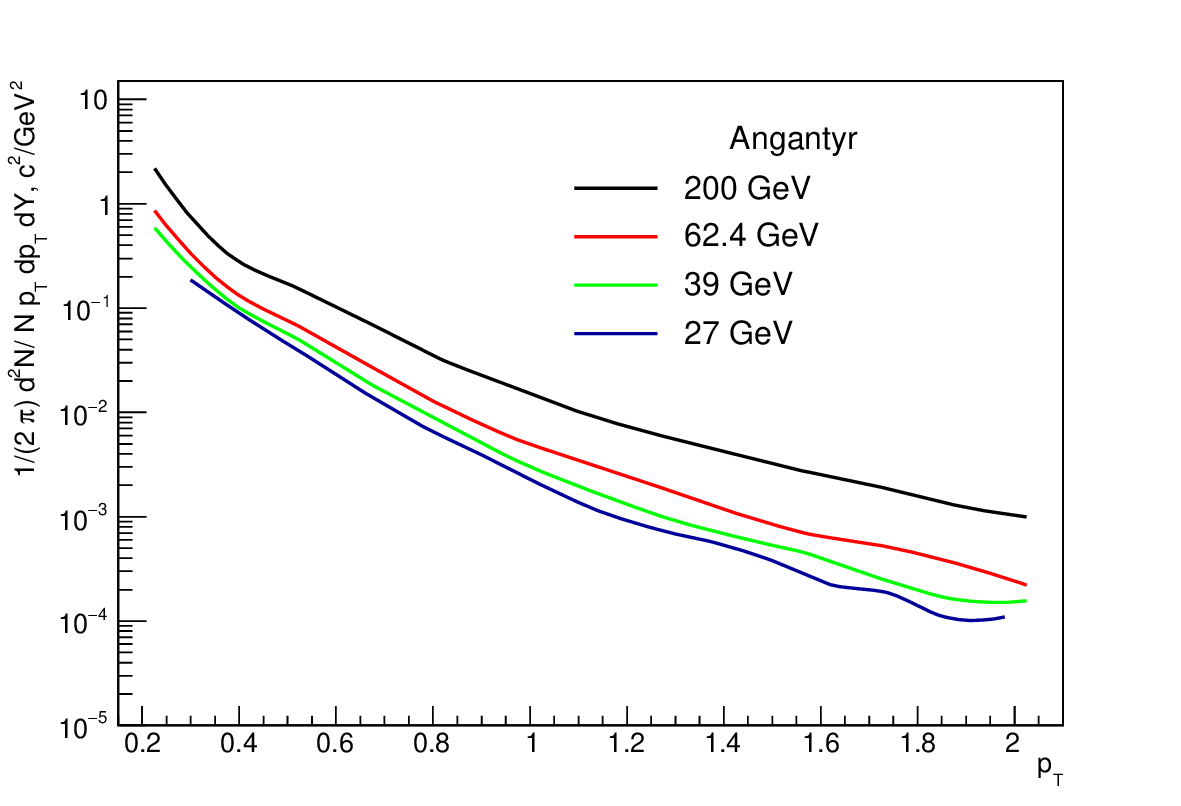}
\caption{\label{fig:4} The transverse momentum spectra of prompt photons and rest of the decays at different beam energies. }
\end{figure}

Multiple scattering or Multiparton interactions (MPI) are generated in PYTHIA8 based on leading order QCD cross section and regulated by introducing screening parameter $p_{T0}$ \cite{Helenius:2019gbd} i.e.
\begin{eqnarray}
\frac{d\sigma}{dp^{2}_T} \sim \frac{\alpha_{s}^{2} p^{2}_T}{p^{4}_T}
\rightarrow \alpha_{s}^{2} \frac{p^{2}_{T0}+p^{2}_T}{(p^{2}_{T0}+p^{2}_T)^2}
\label{eq:1}
\end{eqnarray}
The screening parameter $p_{T0}$ is assumed to be dependent on beam energy and a tunable parameter $p^{ref}_{T0}$ is used to describe such dependence i.e. 
\begin{eqnarray}
p_{T0}(\sqrt{s})= p^{ref}_{T0}(\frac{\sqrt{s}}{\sqrt{s_{ref}}})^p
\label{eq:2}
\end{eqnarray}
Where $\sqrt{s_{ref}}$ is the reference center of mass(c.m) energy. The multiplicity of the produced particles is sensitive to $p^{ref}_{T0}$ \cite{Weber:2018ddv,Helenius:2019gbd}. A lower value of $p^{ref}_{T0}$ can lead to high multiplicity. The value of $p^{ref}_{T0}$ is tuned for various collision systems and energies tuned by matching the charged particle multiplicity $\frac{dN_{ch}}{d\eta}$ with the experimental data within uncertainties. The values of $p^{ref}_{T0}$ for different collision systems and the corresponding $\frac{dN_{ch}}{d\eta}$ can be found in Table 1. The final hadronization is done using the LUND string fragmentation model and the probability of creating hadrons from the initial state of partons is computed using LUND area law \cite{Andersson:2001yu}. The channels for prompt photon production used for the work are $gg\rightarrow q\gamma, q\bar{q} \rightarrow g\gamma $ $, gg\rightarrow g\gamma $ $, q\bar{q} \rightarrow \gamma \gamma $ $, $  and $gg \rightarrow \gamma \gamma$, the matrix elements are from the corresponding papers \cite{Costantini:1971cj,Berger:1983yi,Dicus:1987fk}. These processes are also included in the soft part of the model and their production is controlled using $p^{0}_T$.

Besides the hard processes, the Quantum Electrodynamics (QED) processes during Parton showers are a major source of photons. The charged particles produce photons in the initial state showers by Bremmstraulung. QED shower in PYTHIA8 incorporates a fully coherent multipole treatment of photon radiation off systems of charged fermions, vectors, and scalar particles,
as well as photon splittings to pairs of charged fermions \cite{Skands:2020lkd,Kleiss:2017iir,vonWeizsacker:1934nji,Williams:1934ad,Ohl:1996fi}.

 It should be noted that the Angantyr model does not assume a hot thermalized medium and does not calculate the collective effects. Hence, the model can be used as a good tool to distinguish between collective and non-collective effects in heavy ion collision systems. 
\section{Results}
 We have generated 2x$10^6$ events for each collision system with SoftQCD:all = on and the tuned values of $p^{ref}_{T0}$. The events are divided into different centrality classes based on $\Sigma E_T$ of the charged hadrons in the rapidity range $-0.35<y<0.35$ to match the kinematic range of the PHENIX detector \cite{Singh:2022ijk}.
 
\subsection{Different sources of photons}

The photon spectra can be distinguished in two parts: 1. Direct photons and 2. Decay photons. In Fig.1 we have shown the photon spectra for central Au-Au collisions at 200 GeV, the direct and decay photons are shown separately for better visualization. It can be seen that the yield from the decay is much higher in magnitude than that of the prompt photons. The major contributor to the decay photon yield is the neutral pion ($\pi^{0}$) followed by $\eta$ meson, while the contributions from $\eta'$ and $\omega$ are small but still greater than the prompt photons. The other particles such as $\Lambda$, $\phi$, etc. which are not removed from the data in the experiments have a little contribution compared to other sources but they exceed the prompt photons at low $p_{T}$ and their contributions should be considered for a more accurate identification of direct photons.

 The prompt photons are a combination of primary hard scattering, Multiparton interactions (MPI), and parton showers. The Angantyr simulations with all processes enabled, with MPI=off, and with both MPI=off and ISR=off are shown in Fig 1. This suggests that apart from primary hard scattering photons are produced in the secondary interactions and Bremmstrahulang-like processes. It can be seen that most of the low $p_T$ photons are produced from the initial state radiation (ISR) in the early stage of collision. The simulation without MPI has an invariant yield similar to that of the "MPI on" case. This suggests that MPI does not significantly affect photon production at high $p_T$. However, the MPIs can produce photons in the subsequent hard processes (which is absent in HardQCD simulations), enhancing the low $p_T$ spectra. Regardless of the configuration the model matches with the experimental data \cite{STAR:2016use} at high $p_T$ ($>5$ GeV) within error bars, this is expected as the high $p_T$ photons are due to the contribution from primary hard scatterings. The prompt photon yield in Angantyr is lower than a typical scaled p-p collision yield. This is due to sub-collisions being separated into absorptive, wounded target, wounded projectile, double diffractive, and elastic interactions. A clear depiction can be seen in Fig 1. 

 The experiments often estimate the contribution from all hadronic decays by simulating the ratio of all decay photons to the contribution from neutral pions ($\gamma ^{hadrons}/\gamma ^{\pi ^0}$). The Angantyr results without rescattering are similar to PHENIX Exodus \cite{PHENIX:2022rsx} simulations for Au-Au collisions at 200 GeV, 62.4 GeV, and 39 GeV. The ratio is almost independent of beam energy but the smaller systems like d-Au and He-Au tend to have a lower ($\gamma ^{hadrons}/\gamma ^{\pi ^0}$) ratio than larger systems like Au-Au at 200 GeV. The hadronic rescatterings can lower the ($\gamma ^{hadrons}/\gamma ^{\pi ^0}$) ratios for all energies. A brief description of rescatterings is given below.

\begin{figure}[h]
\includegraphics[width=.5\textwidth]{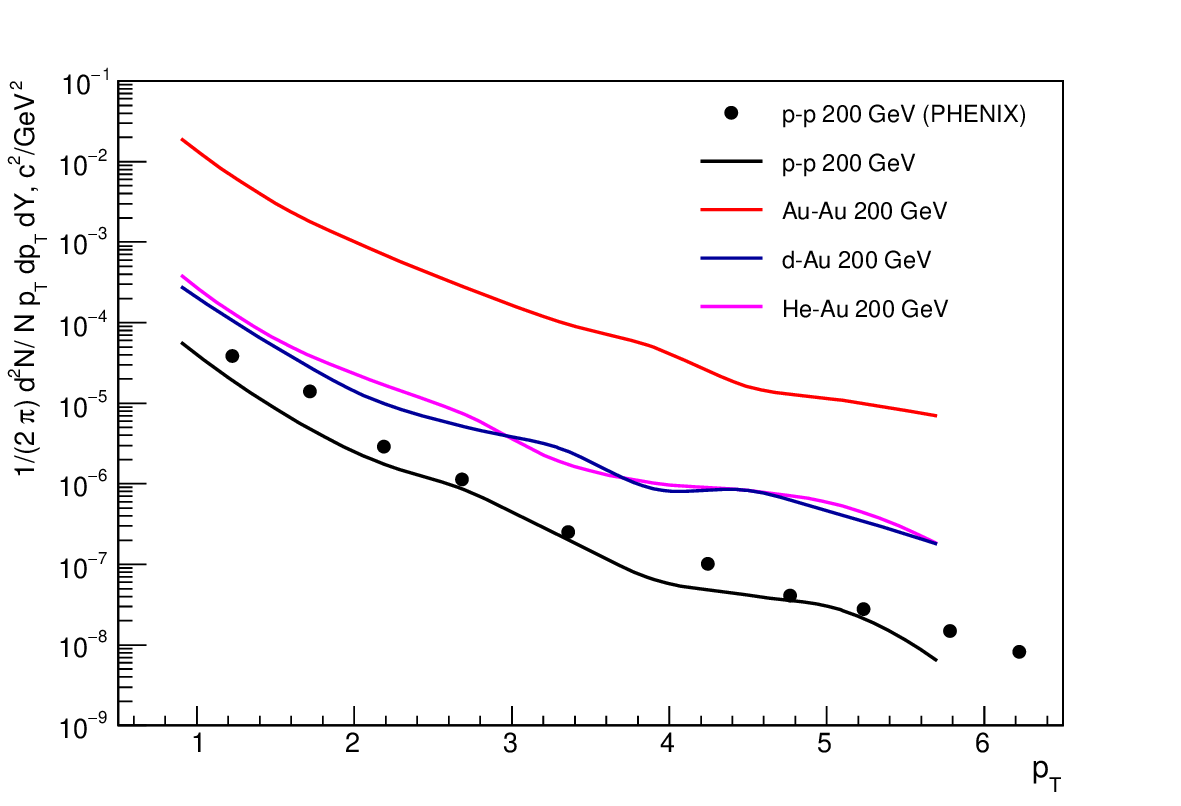}
\caption{\label{fig:5} The $p_T$ spectra of prompt photons for most central collisions (0-20)\% for different collision systems at 200 GeV. The p-p collision data is from the corresponding paper \cite{PHENIX:2022rsx}. }
\end{figure}

\begin{figure}[h]
\includegraphics[width=.5\textwidth]{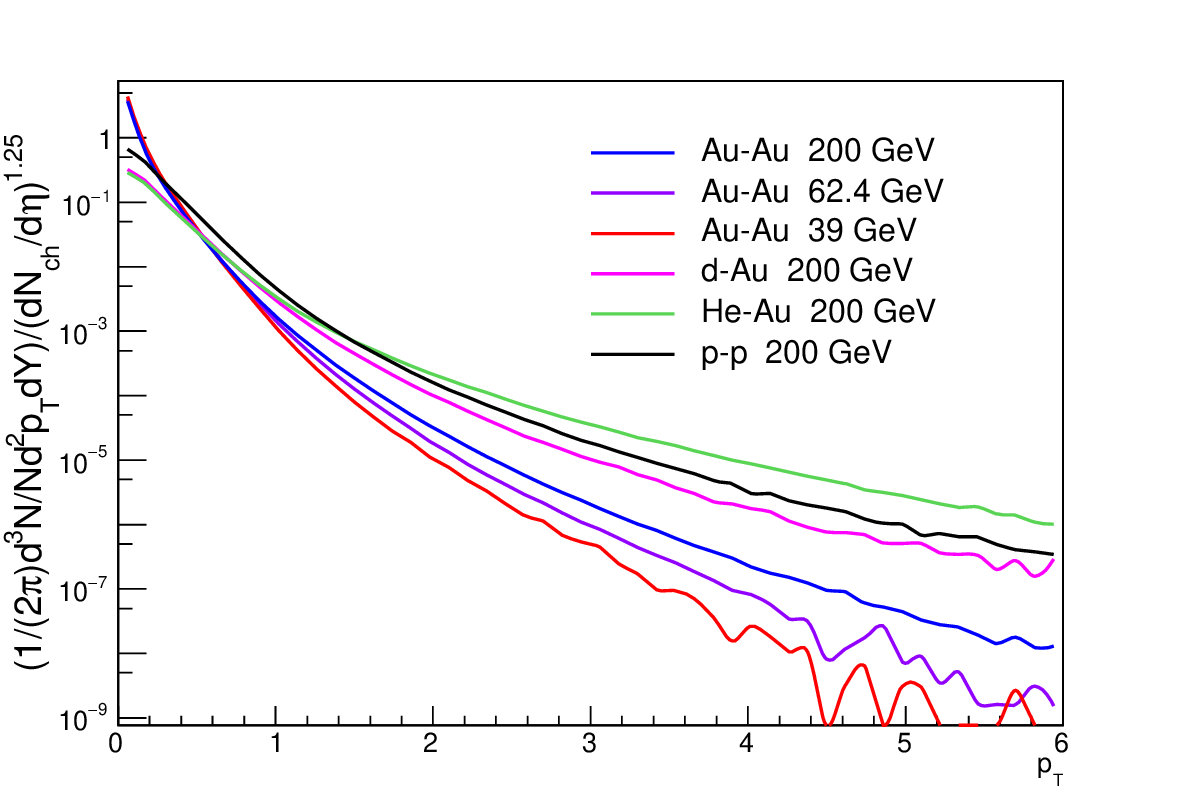}
\caption{\label{fig:5} The $\frac{dN_{ch}}{d\eta}^{1.25}$ scaled $p_T$ spectra of decay photons for most central collisions (0-20)\% for different beam energies. The p-p results are for minimum bias collisions. }
\end{figure}

 The processes after the hadronisation like the hadronic rescatterings also produce photons. The rescattering is significantly high in a heavy-ion event compared to a typical p-p event due to its high multiplicity, hence it should be considered while studying the final state particles. The physics processes are similar to low-energy QCD processes \cite{Bierlich:2022pfr}. The hard QCD processes and MPIs are not included as the energy of these collisions will be way below 10 GeV. A detailed rescattering framework used in this work can be found in the corresponding paper\cite{Bierlich:2021poz}. The rescattering starts early in Angantyr due to the absence of QGP, hence its effects in Angantyr are more pronounced than the QGP-based models like UrQMD and SMASH which use rescattering as afterburner\cite{Bass:1998ca,SMASH:2016zqf}. Considering the low energy scales of the processes the prompt photon production is unlikely but the same can not be told about the decay photons. Many exotic resonances are formed during the rescatterings and decay into photons (these photons are often considered direct but we will refer to them as decay photons due to the nature of their origin). The major contributors are particles like the $f_0(600)$ meson which is produced abundantly in rescattering due to its low mass. In a rescattering scenario, the contribution from the resonances is greater than $\eta'$ and $\omega$ combined. The contributions from the resonances are distinct at low $p_T$, after 1.5 GeV their contribution vanishes. It is noteworthy that the thermal photons dominate the low $p_T$ region and these resonances' effect can significantly impact their identification. A clear depiction can be seen in Fig 3. 
 
 The prompt photons can also be produced from the interactions between mesons and baryons as shown in the corresponding paper \cite{Linnyk:2015tha}. Some of the major processes include $\pi+\pi \rightarrow \rho +\gamma$, $\pi+\rho \rightarrow \pi+\gamma$, and $V + N \rightarrow \gamma + N$. Where V is a vector meson and N is a nucleon. However, the contributions from these processes are quite small compared to the other sources. The contributions from $\pi \pi$, $\pi \rho$, and $VA$ processes from Linnyk et al. \cite{Linnyk:2015tha} are shown in Fig 3 for comparison.

The direct photon spectra take the following empirical form at low $p_T$ (around 0.4 to 1.3 GeV/c) \cite{PHENIX:2022qfp}:
\begin{eqnarray}
\frac{d^2 N}{2\pi dp_Tdy} \approx e^{-(p_T/T_{eff})}
\label{eq:3}
\end{eqnarray}
 In a rescattering scenario, the extra decay photons can affect the $T_{eff}$ if not removed from the data. We removed the prompt photons and the photons produced in rescattering from the experimental data and fitted the excess photon spectra slope with the exponential function described above to extract the effective temperature (inverse slope $T_{eff}$) of the medium formed. The best match was observed for $T_{eff}$ = 0.167 GeV/c for central Au-Au collisions at 200 GeV. A similar fitting was performed in the corresponding paper \cite{PHENIX:2022qfp}. However, the experimental data (which still has the background decay photons ) was used for the fitting, the previous estimate of $T_{eff}$ was about 0.178 GeV/c for central collisions at 200 GeV which is higher than the current values obtained from the fit.  A clear depiction of excess photons and the exponential fits can be found in Fig 4. In the presence of a thermalized medium, the effect of rescattering is expected to be lower than Angantyr which will lead to a higher $T_{eff}$ than the one extracted in this work. The $T_{eff}$ should still be lower than the previous estimates which calculated $T_{eff}$ by fitting the data without considering rescattering.

In Fig 5, we have shown the transverse momentum spectra of prompt photons in central Au-Au collisions at 200 GeV, 62.4 GeV, 39 GeV, and 27 GeV. We have included the decay photons which are not removed from the data. This was done to show the effect of these decays at low $p_T$. 
\subsection{Photon production in smaller systems}

 The presence of thermal photons suggests that a hot dense state (QGP) is formed in the Au-Au collisions at 200 GeV, 62.4 GeV, and 39 GeV which emits the extra photons \cite{PHENIX:2022rsx}. A major difference between the p-p collision system and the Au-Au system is the existence of thermal photons. We have simulated p-p collisions at 200 GeV using PYTHIA8. The simulations well describe both the multiplicity and $p_T$ spectra of direct photons suggesting that no QGP is formed in p-p collisions at 200 GeV. Smaller systems like d-Au, and 3He-Au are not guaranteed to form QGP at 200 GeV. An excess of direct photons from these systems would imply the existence of a hot thermalized medium. The prompt photon spectra for central (0-20)  \%  d-Au, and 3He-Au collisions can be found in Fig 6. The prompt photon yield increases steadily as the collision system gets larger. A similar trend is also seen for the decay photons. As these collisions are a part of the future RHIC experiments these results can be instrumental in identifying the thermal photons. They can justify or deny the existence of QGP in these collisions.

 The direct photon spectra scales with $\frac{dN_{ch}}{d\eta}^{1.25}$. The $\frac{dN_{ch}}{d\eta}^{1.25}$ is roughly proportional to $N_{part}$, while it does not saturate at large beam energy like $N_{part}$. It increases rapidly with beam energy and gives a good estimate of system size at hadronization. It also has a fairly simple relationship with $N_{coll}$ i.e. $ N_{coll} = (1/\sqrt{SY}) \frac{dN_{ch}}{d\eta}^{1.25}$, where SY is the specific yield which varies logarithmically with beam energy \cite{PHENIX:2018for}.

In Fig 7, we have shown the $\frac{dN_{ch}}{d\eta}^{1.25}$ scaled $p_T$ spectra of decay photons for central Au-Au collisions at 200 GeV, 62.4 GeV, 39 GeV, 27 GeV, and for p-p, 3He-Au, and d-Au collisions at 200 GeV. The decay photons show a scaling behavior similar to the ones observed for direct photons in experiments. The scaling continues up to 1 GeV, after 1 GeV it becomes beam energy dependent. It is expected as the increase in multiplicity increases the number of decays. The hard photons show no such trend. The scaled yield of decay photons decreases with decreasing beam energy. At a given collision energy the scaled yield of decay photons in smaller systems like p-p, 3He-Au, etc is greater than the larger systems like Au-Au. This suggests that the smaller systems produce more background photons for a given size.

\section{Summary}

In the current work, we have simulated the transverse momentum spectra of direct as well as decay photons in Au-Au, d-Au, and 3He-Au collisions using the PYTHIA 8/ Angantyr. The hadronic decays produce most of the photons in an Au-Au collision. The main contribution come from the $\pi^0$ meson followed by $\eta$, $\eta^,$ and $\omega$. The contributions from the rest of the decays are fairly small. The direct photons produced in our simulations are a combination of primary hard scatterings, multi-parton interactions, and initial state radiations. Most of the low $p_T$ photons are produced in initial state radiation while the high $p_T$ photons are produced in the primary hard scatterings. The Angantyr algorithm divides the sub-collisions into absorptive, wounded target, wounded projectile, double diffractive, and elastic interactions hence the calculations are lower than a $N_{coll}$ scaled p-p collision. The model shows a good match with the experimental data at high $p_T$. 

The processes such as hadronic rescattering can lead to the production of a large number of resonances which can decay into photons (such as $f_0 (600)$). The particle can enhance the decay photons and reduce the $\gamma ^{hadrons}/\gamma ^{\pi ^0}$ ratios which are used to separate the decay photons from the direct photons. In a rescattering scenario, these extra photons affect the background of decay photons, ultimately affecting the precise measurement of thermal photons. We have removed the extra decay photons along with the prompt photons from the data and re-calculated $T_{eff}$ in central Au-Au collision at 200 GeV and 62.4 GeV using the inverse slope method. We found a lower value of $T_{eff}$ than the previously estimated values.

We have predicted the direct photon spectra for the smaller collision systems like p-p, d-Au, and 3He-Au collisions at 200 GeV. These results can be used as backgrounds to precisely identify the thermal photons in these systems. The decay photons in these systems as well Au-Au systems show a similar $\frac{dN_{ch}}{d\eta}^{1.25}$ scaling like the thermal photons at low $p_T$. The scaling breaks at high $p_T$ and becomes beam energy dependent. The Au-Au collisions at 200 GeV have a higher scaled yield than the lower energies. However, the smaller collision systems at 200 GeV tend to have higher scaled yield than the Au-Au collisions.

\section*{Acknowledgements}
BKS sincerely acknowledges financial support from the Institute of Eminence (IoE) BHU Grant number 6031. SRN and GD acknowledges the financial support from the UGC Non-NET fellowship during the research work. 



\end{document}